OPTICS

# Electrochemically controlled metasurfaces with high-contrast switching at visible frequencies

Robin Kaissner[1†], Jianxiong Li[2*†], Wenzheng Lu[3], Xin Li[4], Frank Neubrech[1], Jianfang Wang[3*], Na Liu[1,2*]

**Recently in nanophotonics, a rigorous evolution from passive to active metasurfaces has been witnessed. This advancement not only brings forward interesting physical phenomena but also elicits opportunities for practical applications. However, active metasurfaces operating at visible frequencies often exhibit low performance due to design and fabrication restrictions at the nanoscale. In this work, we demonstrate electrochemically controlled metasurfaces with high intensity contrast, fast switching rate, and excellent reversibility at visible frequencies. We use a conducting polymer, polyaniline (PANI), that can be locally conjugated on preselected gold nanorods to actively control the phase profiles of the metasurfaces. The optical responses of the metasurfaces can be in situ monitored and optimized by controlling the PANI growth of subwavelength dimension during the electrochemical process. We showcase electrochemically controlled anomalous transmission and holography with good switching performance. Such electrochemically powered optical metasurfaces lay a solid basis to develop metasurface devices for real-world optical applications.**

## INTRODUCTION

Active functions are indispensable features of modern optical devices. For instance, in a digital camera, the zoom lens is a key component that dynamically varies the focal length and thus the angle of view. In a display device, such as spatial light modulators, dynamic images are processed with rapid speeds, high contrast, and full reversibility. Nevertheless, these state-of-the-art optical devices are often bulky and difficult to miniaturize. Recently in nanophotonics, a new class of optical elements, called metasurfaces, have been rigorously developed. With subwavelength antennas arranged in two-dimensional planes, metasurfaces have exceptional control over the propagation of light, enabling flat, ultrathin, and highly compact optical systems.

Despite the exciting progress, studies on active metasurfaces still await further research endeavors. So far, metal- and dielectric-based metasurfaces have been continuously improved to exhibit remarkable performance and high efficiencies (*1*–*7*). However, the working elements of the metasurfaces, namely, the metallic or dielectric antennas themselves, are often intrinsically static. Modifying the antennas to yield active responses, for instance, by chemical transformations, is one effective solution (*8*–*12*). However, this usually comes at the price of weak optical signals and incomplete reversibility. Alternatively, metasurfaces can be endowed with active responses indirectly using functional materials as surrounding media (*13*–*19*). Such material candidates are abundant. A fundamental selection criterion is that the optical properties of the antennas, including the resonance position, linewidth, and intensity, can be drastically altered along with the electronic (*20*–*22*), mechanical (*23*–*25*), chemical, or structural changes of the surrounding media upon external stimuli (*26*).

To this end, a number of active metasurfaces have been implemented by using functional materials as substrates or hosting matrices for antennas (*27*–*32*). However, such metasurfaces in general encounter poor device control with low-intensity modulations, because the optical properties of all the antennas are altered simultaneously, resulting in limited tuning flexibilities. In particular, this strategy is not effectively applicable to metasurfaces based on the Pancharatnam-Berry (PB) phase, which is a widely applied design blueprint for metasurfaces. The PB phase is achieved by simply controlling the in-plane orientations of optical antennas. It does not depend on the specific antenna geometry or wavelength, thus enabling broadband performance (*29*, *33*, *34*). In turn, due to the broadband nature and thus wavelength insensitivity, the resonance position tuning that active media generally impart to optical antennas is not directly exploitable.

In this work, we demonstrate electrochemically controlled metasurfaces with high-contrast switching at visible frequencies. The metasurfaces are designed based on the PB phase with preselected antennas locally conjugated with a conducting polymer, polyaniline (PANI), of subwavelength dimension. The optical responses of the metasurfaces can be in situ monitored and optimized by controlling the PANI thickness during its electrochemical growth. The operation of the metasurfaces is electrochemically powered, showing excellent performance with high intensity contrast up to 860:1, fast switching rate around 35 ms, and notable reversibility over 100 switching cycles without substantial degradations.

## RESULTS

### Working principle of the electrochemically controlled metasurface

Among a variety of conducting polymers, PANI is particularly attractive because of its high stability, low cost, and facile synthesis (*35*). When electrochemically switched between the oxidized form [emeraldine state (ES)] and reduced form [leucoemeraldine state

[1]2nd Physics Institute, University of Stuttgart, Pfaffenwaldring 57, 70569 Stuttgart, Germany. [2]Max Planck Institute for Solid State Research, Heisenbergstrasse 1, 70569 Stuttgart, Germany. [3]Department of Physics, The Chinese University of Hong Kong, Shatin, Hong Kong SAR, China. [4]School of Optics and Photonics, Beijing Institute of Technology, Beijing, 100081, China.
*Corresponding author. Email: jxli@is.mpg.de (J.L.); jfwang@phy.cuhk.edu.hk (J.W.); na.liu@pi2.uni-stuttgart.de (N.L.)
†These authors contributed equally to this work.

(LS)] as shown in Fig. 1A, PANI undergoes large variations of its refractive index in dependence on the applied voltage, especially in the imaginary part (*36*, *37*). More specifically, at the ES, PANI exhibits strong absorption, whereas at the LS, PANI has almost no absorption in the visible and near-infrared wavelength regimes (*38*). Electrochemical switching of PANI features ultrafast speed around 1 μs and can be operated more than $10^5$ times without any degradation (*39*). PANI is thus an ideal functional material to endow metasurface devices with active optical responses (*39*–*44*).

The design principle of the electrochemically controlled metasurface is shown in Fig. 1B. The metasurface contains two sets of pixels that are alternatively distributed in odd and even rows on an indium tin oxide (ITO)–coated quartz substrate. One set consists of static pixels, which are gold nanorods (200 nm by 80 nm by 50 nm) embedded in polymethyl methacrylate (PMMA; light pink) in the odd rows. The total height of the PMMA-covered rows is $h_1$, and the refractive index of PMMA is $n_1 + ik_1$. In the visible wavelength regime, $n_1 = 1.5$ and $k_1 \approx 0$ (*45*). The other set consists of active pixels, which are gold nanorods locally conjugated with PANI (light blue) in the even rows. The total height of the PANI-covered rows is $h_2$, and the refractive index of PANI is $n_2 + ik_2$. Specifically, the thickness of PANI coated on the gold nanorod surface is defined as $t_{PANI}$, i.e., $h_2 = 50\text{ nm} + t_{PANI}$. The gold nanorods are arranged on the substrate based on the PB phase, which is solely determined by the nanorod orientations for wavefront shaping of circularly polarized light (CPL).

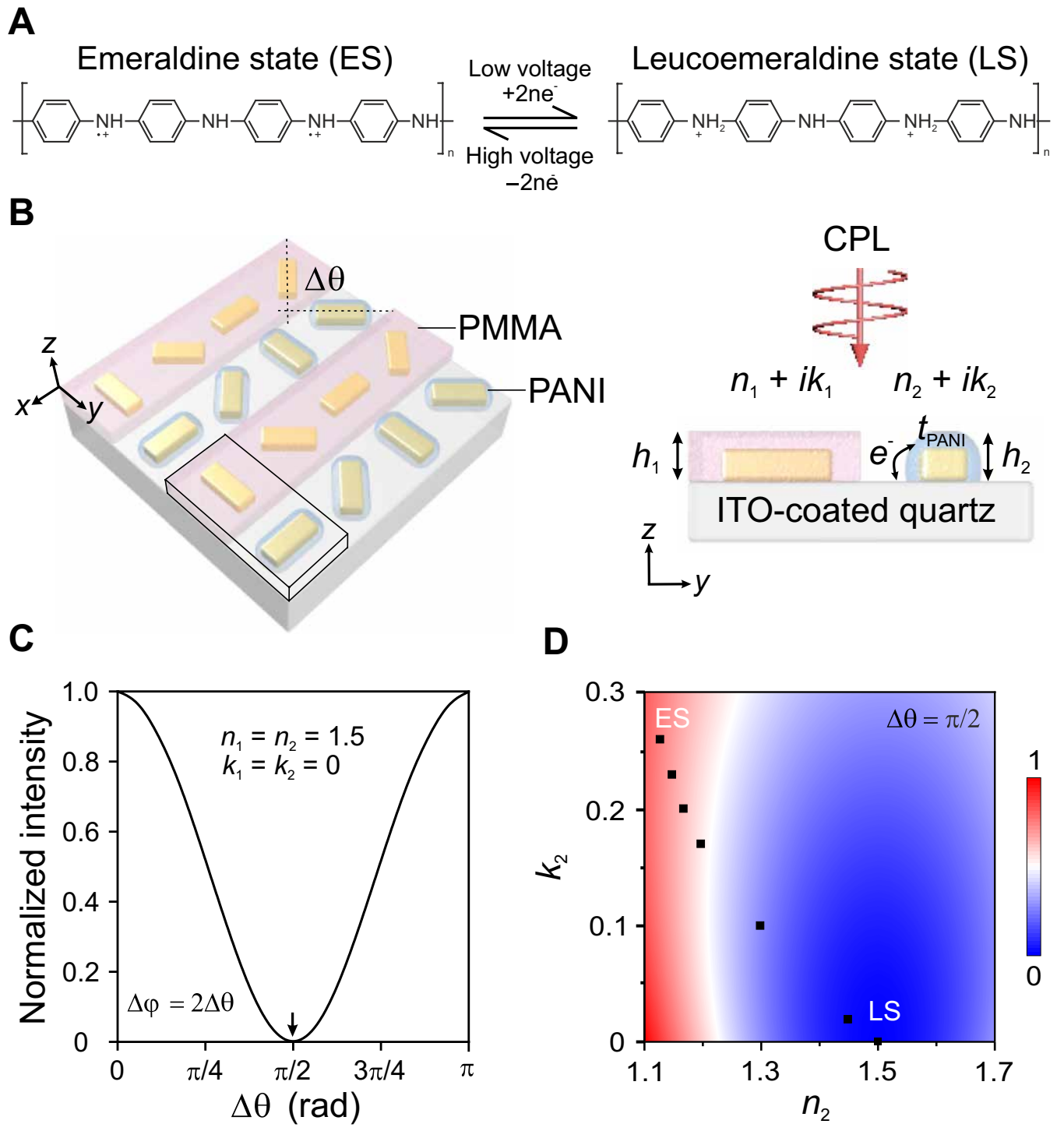

**Fig. 1. Working principle of the electrochemically controlled metasurface.** (**A**) Chemical structures of PANI at the ES and LS controlled by the applied voltage. (**B**) Schematic of the electrochemically controlled metasurface. Two sets of gold nanorods (200 nm by 80 nm by 50 nm) are arranged in alternating rows on top of an ITO-coated quartz substrate. One set is entirely covered by PMMA (height $h_1$), whereas the gold nanorods in the other set are locally conjugated with PANI (thickness $t_{PANI}$, $h_2 = t_{PANI} + 50$ nm). The complex refractive indices of PMMA and PANI are $n_1 + ik_1$ and $n_2 + ik_2$, respectively. (**C**) Normalized intensity of the anomalous transmission in dependence on the nanorod angle difference Δθ simulated at an operating wavelength of 633 nm. Here, $n_1 = n_2 = 1.5$, $k_1 = k_2 = 0$, and $h_1 = h_2 = 100$ nm. (**D**) Simulated anomalous transmission as a function of $n_2$ and $k_2$, when $n_1 = 1.5$, $k_1 = 0$, and $\Delta\theta = \pi/2$. The black squares indicate the complex refractive indices ($n_2$, $k_2$) of PANI at different applied voltages.

A phase difference factor Δφ is introduced between the light waves scattered by the odd and even rows. The angle between two adjacent gold nanorods in the neighboring odd and even rows is defined as Δθ (see Fig. 1B). To elucidate the concept, a metasurface designed for active anomalous transmission is investigated theoretically and experimentally. In the numerical simulations, the metasurface is first entirely covered by a homogeneous PMMA layer ($n_1 = 1.5$) and the operating wavelength is 633 nm. The simulated anomalous transmission as a function of Δθ is shown in Fig. 1C. When $\Delta\theta = \pi/2$ (indicated by the arrow), Δφ is π according to the PB phase. There is no light output from the metasurface due to destructive interference between the scattered light from the neighboring rows. In contrast, when Δθ = 0 or π (Δφ = 0 or 2π), the anomalous transmission of the metasurface reaches the maximum intensity in the far field due to constructive interference.

With PMMA coating and local PANI conjugation in alternating rows as shown in Fig. 1B, the light output of the metasurface can be actively modulated through electrochemical control of PANI conjugated on the preselected gold nanorods. Figure 1D presents the simulated anomalous transmission intensity of the metasurface in dependence on the refractive index tuning of PANI, while the refractive index of PMMA is fixed at $n_1 = 1.5$. The refractive indices ($n_2$, $k_2$) of PANI at different applied voltages are retrieved from the experimental optical spectra (see fig. S1) and the corresponding data points are marked using black squares in Fig. 1D. $h_1$ and $h_2$ are set as 100 nm in the simulations. As shown in Fig. 1D, when PANI is close to its reduced form (LS) with almost no absorption ($k_2 \approx 0$), the anomalous transmission becomes very weak. In particular, when $n_2 = 1.5$, the light intensity approaches zero. This is evident because when $n_1 = n_2$, it is equivalent to the case of a metasurface homogeneously covered by PMMA (see Fig. 1C). The intensities of the scattered light from the neighboring rows are equal and Δφ remains to be π. In contrast, when $k_2$ increases and $n_2$ decreases, the anomalous transmission gradually gains strength and achieves an intensity maximum at the oxidized form (ES) as shown in Fig. 1D. In this case, PANI is highly absorptive and the scattered light from the PANI-conjugated rows is suppressed. The metasurface output is thus mainly determined by the PMMA-covered rows. Meanwhile, $n_2$ substantially differs from $n_1$, resulting in a resonance shift between the antennas in the neighboring rows. As a result, Δφ largely deviates from π, giving rise to strong light output from the metasurface. In other words, light switching with high intensity contrast can be achieved by electrochemical tuning of the locally PANI-conjugated rows on the metasurface (see fig. S2).

## Electrochemical growth of PANI on preselected gold nanorods

To validate our design concept, an electrochemically controlled metasurface is experimentally implemented for active switching of anomalous transmission. It is named scheme A to compare with a control metasurface (scheme B), which will be discussed later. The metasurface comprises gold nanorods of four phase levels fabricated by electron-beam lithography (EBL) on an ITO-coated quartz substrate. The gold nanorods are 200 nm by 80 nm by 50 nm in dimension. The odd rows are covered by PMMA ($h_1$ = 100 nm) through a double-layer EBL process with the help of high-resolution alignment markers. Local growth of PANI on the gold nanorods in the even rows is carried out by electrochemical polymerization of the metasurface sample in an aqueous electrolyte, containing 2 M $HNO_3$ and 0.1 M aniline. As shown in Fig. 2A, a Ag/AgCl electrode and a Pt wire serve as reference and counter electrodes, respectively. The ITO functions as a working electrode. All the electrochemical potentials described in the experiments are measured relative to the Ag/AgCl reference electrode. To locally grow PANI on the gold nanorods in the even rows, cyclic voltammetry is carried out at a cycling rate of 25 mV/s within a potential range from −0.2 to 0.8 V (see Fig. 2B). PANI preferentially grows on the gold nanorods in the even rows, because PMMA blocks the contact between the electrolyte and the gold nanorods in the odd rows (*46*). During the PANI growth, the intensity of the anomalously transmitted light from the metasurface is monitored in real time. The incident light is right-handed circularly polarized (RCP). The anomalously transmitted light, which is left-handed circularly polarized (LCP), is detected at an off-axis angle of 32° (see fig. S3). The corresponding scanning electron microscopy (SEM) image of the sample in Fig. 2C confirms that PANI is locally conjugated only around the gold nanorods in one set of the rows, whereas PMMA entirely covers the other set of the rows. The PANI thickness coated on the gold nanorods is approximately $t_{PANI}$ = 50 nm (i.e., $h_2$ = 100 nm), obtained from the atomic force microscopy measurement (see Fig. 2D). As shown in Fig. 2E, by increasing the coating cycle number and thus the PANI thickness, the intensity contrast of the output anomalous transmission becomes more and more distinct, when driving between low and high voltages. The intensity minima and maxima occur at −0.2 and 0.6 V in the different coating cycles, respectively. At a coating cycle number of 36 (see the red circle), the light intensity approaches zero. This indicates that the condition for $\Delta\varphi = \pi$ has been fulfilled and the optimal $t_{PANI}$ is reached (see fig. S4). The intensity contrast defined as the ratio between the maximum and minimum intensities at the two states, which is a common measure for specification of the contrast of electronic visual display devices, is as high as 860:1.

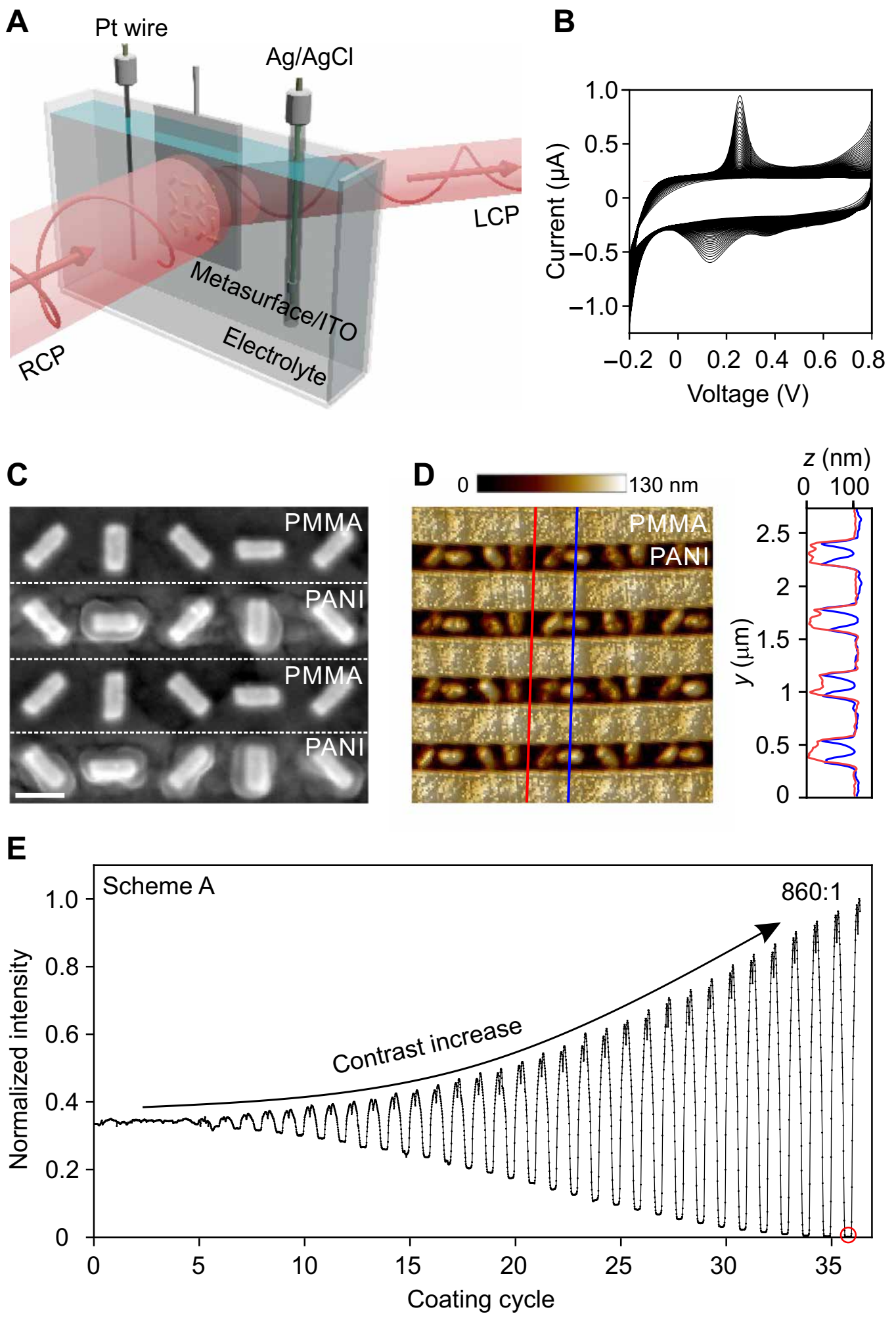

**Fig. 2. In situ optimization of the metasurface performance.** (**A**) Schematic of the experimental setup. The metasurface on ITO (working electrode) is immersed into an electrolyte in a glass cell along with a Pt wire (counter electrode) and a Ag/AgCl reference electrode. Right-handed circularly polarized (RCP) light impinges on the sample at normal incidence, and the anomalous transmission intensity is recorded. (**B**) Cyclic voltammetry diagram for the electrochemical deposition of PANI on the metasurface sample. A potential range from −0.2 to 0.8 V and a scan speed of 25 mV/s are used. (**C**) SEM image of the metasurface with the optimized PANI thickness. Scale bar, 200 nm. (**D**) Atomic force microscopy image of the metasurface and selected height profiles. The measured thickness of PANI coated on the gold nanorods, $t_{PANI}$, is approximately 50 nm (i.e., $h_2$ = 100 nm). (**E**) In situ recorded normalized intensity of the anomalous transmission during the PANI growth (scheme A). The electrochemical process is halted when the intensity reaches the minimum (coating cycle 36, red circle). The intensity contrast defined as the ratio between the maximum and minimum intensities is as high as 860:1.

## Switching performance of the electrochemically controlled metasurface

At this optimal $t_{PANI}$, electrochemical switching of the anomalous transmission is performed. A monomer-free solution containing 0.5 M $HNO_3$ is used as the electrolyte in this case. Figure 3A presents the intensity of the anomalously transmitted light as a function of the applied voltage. It is evident that the light intensity continuously increases until the applied voltage reaches 0.6 V. This stems from the gradual refractive index changes of PANI through electrochemical tuning, while the refractive index of PMMA is fixed as discussed in Fig. 1D. The switching times τ for rise ("off→on") and fall ("on→off") processes are approximately 48 and 35 ms, respectively, as shown in Fig. 3B. The operation of light switching between the on (0.6 V) and off (−0.2 V) states exhibits no substantial degradations over 100 switching cycles (see Fig. 3C). Such switching characteristics, including rapid switching speed, high light intensity contrast, and excellent reversibility, demonstrate the superior performance of our active metasurface device.

## Control experiment

To corroborate the power of our design concept, a control experiment (scheme B) is carried out. As shown in the inset schematic of Fig. 4A, in this case, all the gold nanorods on the metasurface are conjugated with PANI. The same coating cycle number of 36 is adopted for a direct comparison with scheme A. The light intensity of the anomalous transmission, in situ recorded during the PANI growth, is presented in Fig. 4A. It is noteworthy that during the coating process, the intensity maxima and minima occur at the LS and ES of PANI in scheme B, respectively. It is opposite to the case of scheme A (see Fig. 2E). This is due to the fact that scheme B simply relies on the intrinsic PANI switching between its nonabsorptive and highly absorptive forms, whereas scheme A uses active phase tuning mediated by PANI to achieve metasurface switching. The variations of the maximum intensity during the coating process in Fig. 4A can be attributed to the resonance shifts of the gold nanorods resulting from their local environment changes as PANI grows. The performance difference between schemes A (red) and B (black) is clearly manifested in Fig. 4B, in which the dependence of the intensity contrast on the coating cycle number is plotted. At the same coating cycle number of 36, intensity contrasts of 860:1 and 10:1 are achieved with schemes A and B, respectively. The former is nearly 90 times higher than the latter. To accomplish a higher intensity contrast using scheme B, thicker PANI has to be used (see fig. S5). However, this is at the price of largely increased switching time τ, which is proportional to $t_{PANI}^2$ as experimentally demonstrated in Fig. 4C. The intensity contrast in scheme A decreases notably after 36 coating cycles upon reaching the optimal PANI thickness. Further growth of PANI beyond the optimal thickness leads to the deviation of the phase difference Δφ from π at the off state and thus the intensity contrast drops as observed in Fig. 4B. An overview comparison between schemes A and B can be found in table S1.

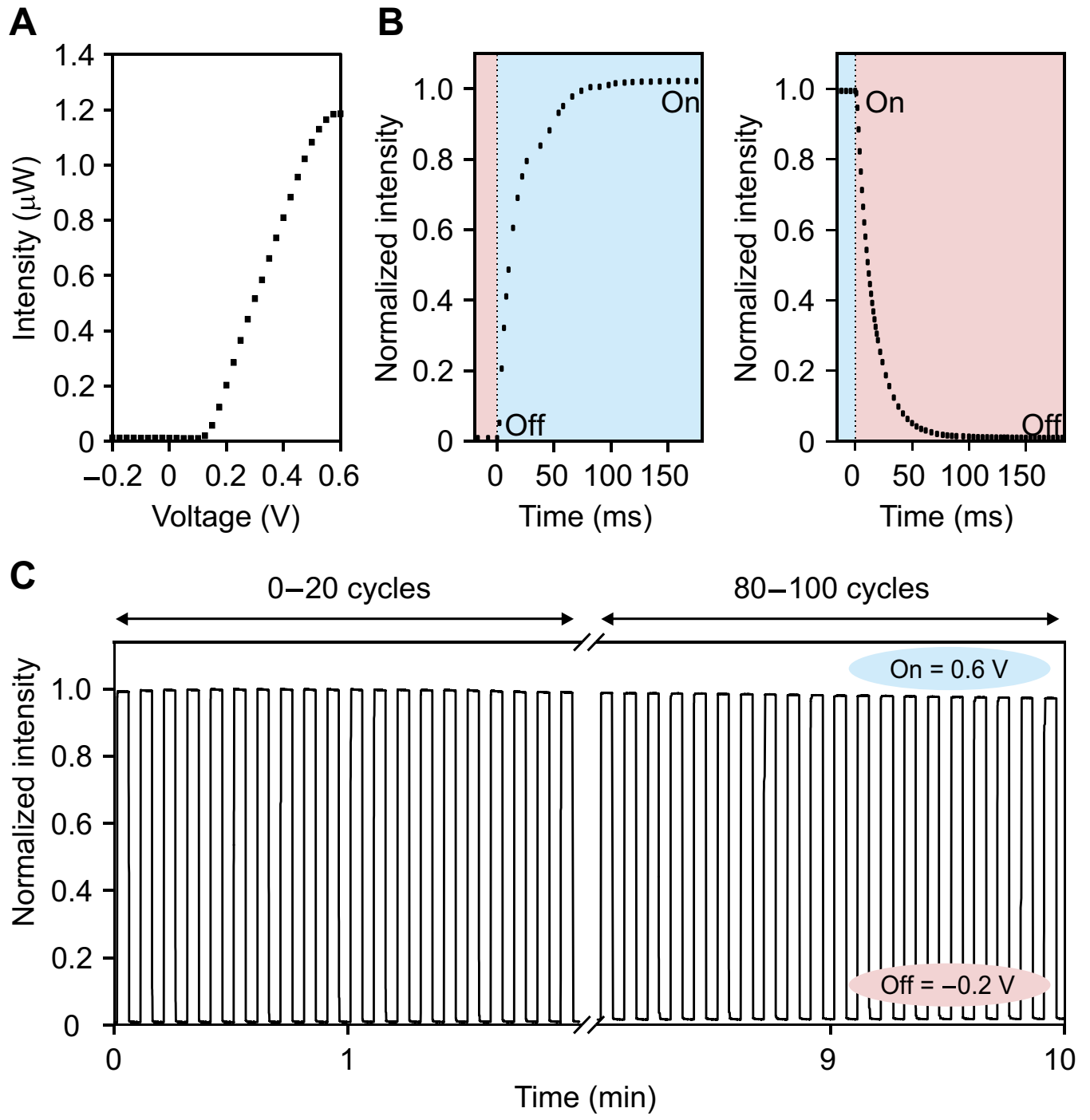

**Fig. 3. Switching performance of the electrochemically controlled metasurface.** (**A**) Intensity of the anomalous transmission in dependence on the applied voltage. (**B**) Switching times for the off→on and on→off processes. The rise time is approximately 48 ms (left) and the fall time is approximately 35 ms (right). (**C**) Highly reversible switching of the anomalous transmission. No substantial degradations are observed over 100 switching cycles. All intensities are normalized by the intensity maximum achieved at the optimal $t_{PANI}$.

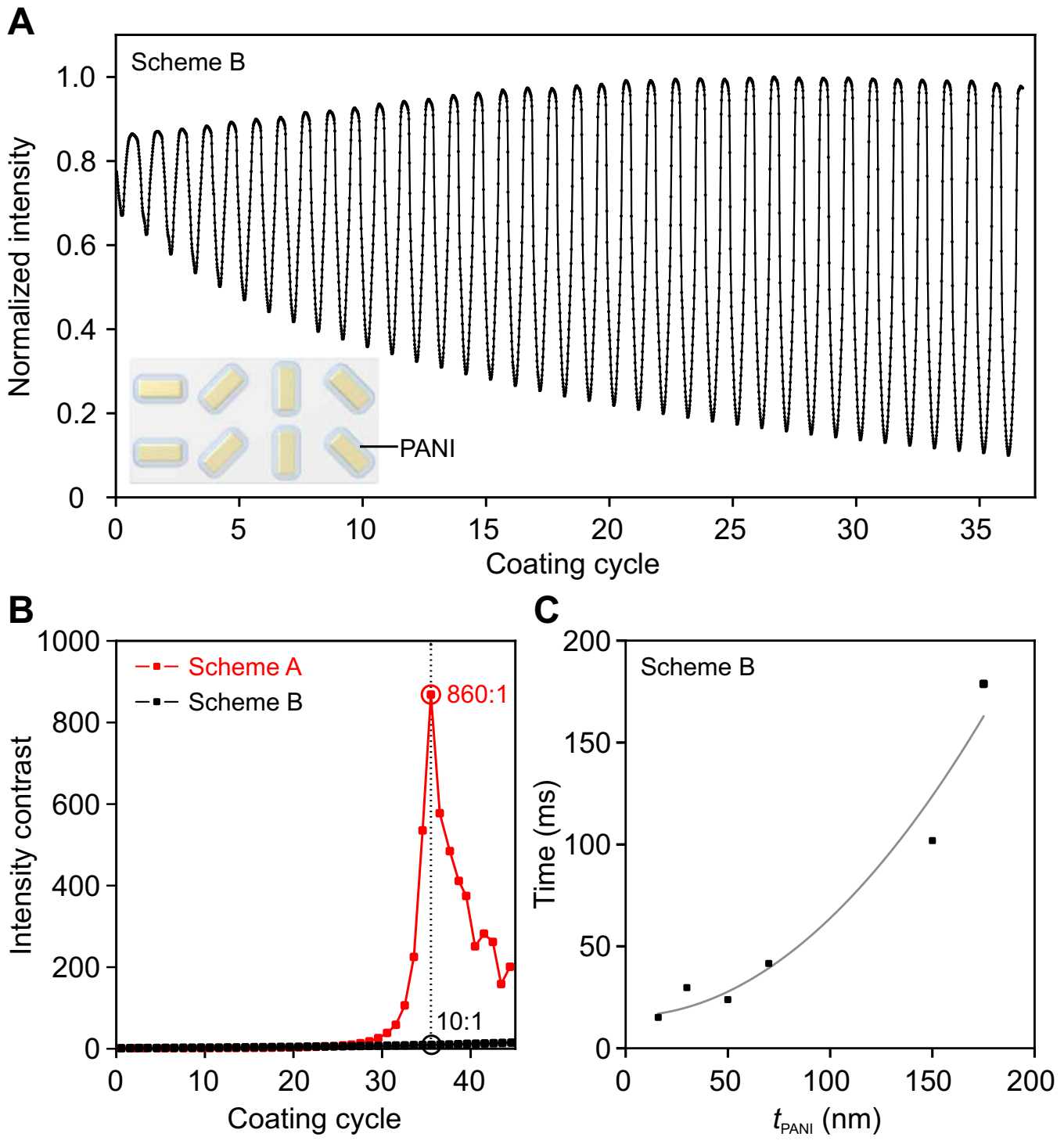

**Fig. 4. Control experiment.** (**A**) In situ recorded normalized intensity of the anomalous transmission during the PANI growth (scheme B). The electrochemical process is halted at a coating cycle number of 36 for a direct comparison with scheme A. Inset, schematic of scheme B, in which all the gold nanorods on the metasurface are conjugated with PANI. The dimensions of the gold nanorods are the same as those in scheme A. (**B**) Light intensity contrasts (ratio between the maximum and minimum intensities) in dependence on the coating cycle for schemes A (red) and B (black), respectively. (**C**) Fall time in dependence on the PANI thickness $t_{PANI}$ in scheme B. The fitted curve follows the relation, $\tau = 0.0048 t_{PANI}^2 + 12.3$. The metasurface switching in scheme B is substantially slowed down for thicker PANI coatings.

### Electrochemically controlled metasurface holography

To demonstrate the generality and versatility of our design concept, next, we apply it for electrochemically controlled addressable metasurface holography. On the basis of the Gerchberg-Saxton algorithm using eight phase levels (*47*), two metasurfaces M1 and M2 are designed to reconstruct holographic patterns with letters "L" and "R," respectively. As shown by the SEM image in Fig. 5A, each metasurface is individually controlled by an ITO electrode. Both the PANI coating and the metasurface switching processes on M1 and M2 are independently operated, enabling electrochemically controlled addressable metasurface holography. The reconstructed holographic patterns are captured by a charge-coupled device camera. Figure 5B shows the snapshots of the addressable holographic patterns L and R independently controlled by electrodes 1 and 2 at their respective on and off states. The intensity profiles of the holographic pattern L along the dashed lines in Fig. 5B at the on and off states are plotted in Fig. 5C. A high intensity contrast between the two states is revealed. Switching demonstration at a different wavelength (520 nm) can be found in fig. S6. It is noteworthy that our design concept can be readily adopted for electrochemically controlled reflective metasurfaces. A reflective metasurface geometry generally contains a back reflector, a transparent dielectric spacer (e.g., $SiO_2$), a conductive layer (e.g., ITO), and metasurface antennas (*6*).

## DISCUSSION

In conclusion, we have demonstrated electrochemically controlled metasurfaces at visible frequencies. The metasurface devices exhibit superior performance with high intensity contrast, fast switching speed, and excellent reversibility. The optical output of the metasurfaces can be in situ optimized by controlling the PANI thickness during its electrochemical growth, offering a facile means to enhance the device performance. With this design concept, the active properties of conducting polymers, which can be produced at low cost and large scale, are readily transferred to optical metasurfaces. Our design concept is generic and can be applied to realize a variety of optical metasurfaces with different active functions, such as beam focusing and zooming and dynamic optical vortices, among others. It can also be easily extended to other frequency regions and adopted for dielectric metasurfaces. Our work will foster a profound research area, in which interesting optical functions and material characteristics are elegantly combined together to create novel nanophotonic devices. With the continuous drive for practical applications, this research area will certainly excite intensive academic and industrial activities.

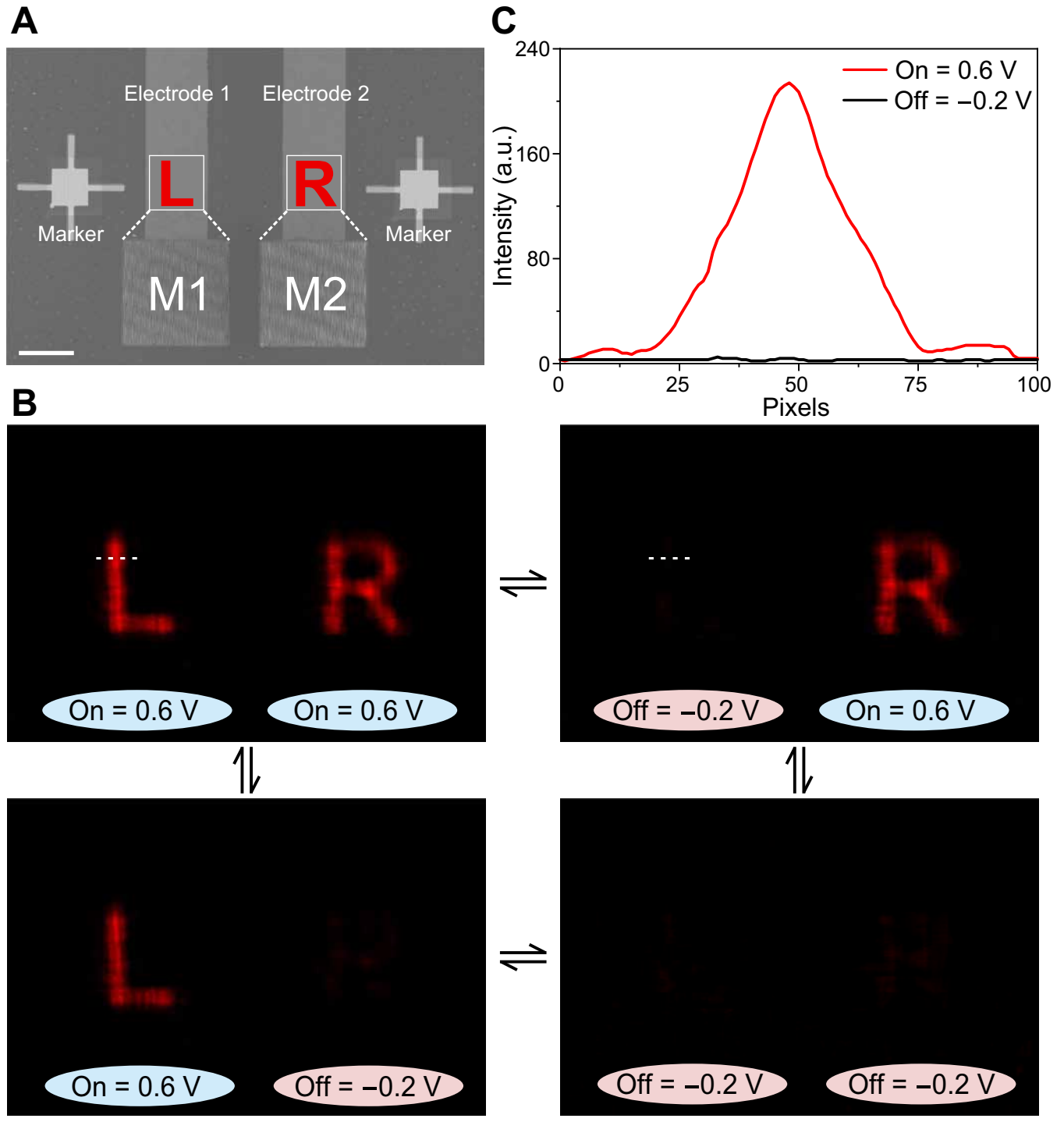

**Fig. 5. Electrochemically controlled addressable metasurface holography.** (**A**) SEM image of the device containing two addressable metasurfaces (M1 and M2) controlled by independent electrodes. M1 and M2 are designed to reconstruct holographic patterns L and R, respectively. The lateral distance between M1 and M2 is 30 μm. Scale bar, 50 μm. (**B**) Experimental results. Holographic patterns L and R are electrochemically switched on and off without cross-talk in an addressable manner. (**C**) Intensity profiles of the holographic pattern L along the dashed lines in (B) at the on and off states, respectively.

## MATERIALS AND METHODS

### Sample fabrication

The metasurface samples in Figs. 2, 3, and 5 were fabricated by two-step EBL using a "Raith eLine" system. In the first EBL step, the gold nanorods and gold alignment markers were defined in PMMA resist (Allresist) on an ITO (20 nm)/$SiO_2$ (1 mm) substrate. A 2-nm Cr adhesion layer and a 50-nm gold layer were deposited by thermal evaporation using a "Leybold Univex" system, followed by liftoff. Subsequently, a 100-nm PMMA layer was spin-coated onto the sample. In a second EBL step, computer-controlled alignment was performed using the alignment markers to precisely place and define the alternating grooves into the PMMA layer. The fabrication of the scheme B metasurface in Fig. 4 was ended after the lift-off step, leaving out the second EBL. Aniline is toxic if ingested or inhaled or through skin contact. Care must be taken when it is used for synthesis.

### Sample characterization

The light beam was generated from a laser diode source (633 nm). A linear polarizer (LP) and quarter–wave plate (QWP) were used to obtain the required incident CP light. A lens and an aperture were used to reshape the light beam. The light beam was incident on the sample, which was immersed in the electrolyte of the electrochemical cell at normal incidence. The intensity of the cross-polarized light was recorded in situ using a power meter (Thorlabs S130C) during the electrochemical deposition of PANI. The holographic images were recorded by a charge-coupled device camera (Thorlabs DCU223C) using an objective (20×). An LP and a QWP were used to filter out the transmitted light with the incident polarization (see fig. S7).

## SUPPLEMENTARY MATERIALS

Supplementary material for this article is available at http://advances.sciencemag.org/cgi/content/full/7/19/eabd9450/DC1

**Acknowledgments:** We thank X. Y. Duan for help with AFM imaging. **Funding:** This project was supported by the European Research Council (ERC Dynamic Nano) grant and the Max Planck Society (Max Planck Fellow). **Author contributions:** R.K., J.L., and N.L. conceived the project. W.L. and J.W. constructed the electrochemical system. X.L. carried out the metasurface hologram designs. R.K. and J.L. performed the experiments and theoretical calculations. F.N. and J.W. made helpful comments to the manuscript. All authors discussed the results, analyzed the data, and commented on the manuscript. **Competing interests:** R.K., J.L., and N.L. are inventors on a patent application related to this work filed by European Patent Office (no. MI 1201-6070-BC-JK , filed 18 December 2020). The authors declare that they have no other competing interests. **Data and materials availability:** All data needed to evaluate the conclusions in the paper are present in the paper and/or the Supplementary Materials. Additional data related to this paper may be requested from the authors.

# Electrochemically controlled metasurfaces with high-contrast switching at visible frequencies

Robin Kaissner, Jianxiong Li, Wenzheng Lu, Xin Li, Frank Neubrech, Jianfang Wang, and Na Liu